\documentclass{ws-ijmpd}
\usepackage[super,compress]{cite}

\def\mnras{MNRAS}
\def\apj{ApJ}
\def\apjl{ApJL}
\def\apjs{ApJS}

\def\aap{A\&A}
\def\aapr{A\&AR}
\def\jcap{JCAP}
\def\nat{Nature}
\def\pasj{PASJ}
\def\physrep{Phys. Rep.}

\def\sovast{Sov. Astron.}

\begin{document}

\markboth{Jorge A. Rueda}
{The white dwarf binary merger model of GRB 170817A}

%
\catchline{}{}{}{}{}
%

\title{THE WHITE DWARF BINARY MERGER MODEL OF GRB 170817A}

\author{J. A. RUEDA,$^{1,2,3,4,7}$ R. RUFFINI,$^{1,2,5,6}$ LIANG LI, $^{1,2,8}$ R. MORADI,$^{1,2,8}$ N. SAHAKYAN,$^{9}$ Y. WANG$^{1,2,8}$}

\address{
$^{1}$
ICRANet, Piazza della Repubblica 10, I-65122 Pescara, Italy\\
%
$^{2}$
ICRA, Dip. di Fisica, Sapienza Universit\`a  di Roma, Piazzale Aldo Moro 5, I-00185 Roma, Italy\\
$^{3}$
ICRANet-Ferrara, Dip. di Fisica e Scienze della Terra, Universit\`a degli Studi di Ferrara, Via Saragat 1, I--44122 Ferrara, Italy\\
$^{4}$
Dip. di Fisica e Scienze della Terra, Universit\`a degli Studi di Ferrara, Via Saragat 1, I--44122 Ferrara, Italy\\
$^{5}$
Universit\'e de Nice Sophia-Antipolis, Grand Ch\^ateau Parc Valrose, Nice, CEDEX 2, France
\\
$^{6}$
INAF, Viale del Parco Mellini 84, 00136 Rome, Italy\\
$^{7}$
INAF, Istituto di Astrofisica e Planetologia Spaziali, Via Fosso del Cavaliere 100, 00133 Rome, Italy\\
 $^{8}$
INAF -- Osservatorio Astronomico d'Abruzzo,Via M. Maggini snc, I-64100, Teramo, Italy\\
$^{9}$
ICRANet-Armenia, Marshall Baghramian Avenue 24a, Yerevan 0019, Republic of Armenia
\\
jorge.rueda@icra.it}

\maketitle

\begin{history}
\received{Day Month Year}
\revised{Day Month Year}
\end{history}

\begin{abstract}
Following the GRB 170817A prompt emission lasting a fraction of a second, $10^8$ s of data in the X-rays, optical, and radio wavelengths have been acquired. We here present a model that fits the spectra, flux, and time variability of all these emissions, based on the thermal and synchrotron cooling of the expanding matter ejected in a binary white dwarf merger. The $10^{-3} M_\odot$ of ejecta, expanding at velocities of $10^9$ cm s$^{-1}$, are powered by the newborn massive, fast rotating, magnetized white dwarf with a mass of $1.3 M_\odot$, a rotation period of $\gtrsim 12$ s, and a dipole magnetic field $\sim 10^{10}$ G, born in the merger of a $1.0$+$0.8 M_\odot$ white dwarf binary. Therefore, the long-lasting mystery of the GRB 170817A nature is solved by the merger of a white dwarf binary that also explains the prompt emission energetics. 
\end{abstract}


\ccode{PACS numbers:}


\section{Introduction}\label{sec:1} 

GRB 170817A is a short gamma-ray burst (GRB) whose prompt emission lasts less than a second was detected by the Gamma-ray Burst Monitor (GBM) onboard the NASA Fermi Gamma-ray Space Satellite,\cite{2017ApJ...848L..12A,2017ApJ...848L..14G} and confirmed by \textit{INTEGRAL}.\cite{2017ApJ...848L..15S} It was subsequently associated with GW170817, a gravitational wave signal reported by the LIGO/Virgo Collaboration about 40 min after the Fermi-GBM circular.\cite{2017ApJ...848L..13A} This initial data was then associated with the optical-infrared-ultraviolet source AT 2017gfo, started to be observed about $12$ h ($\approx 4\times 10^4$ s) after the GRB trigger.\cite{2017Natur.551...67P,2017Natur.551...64A,2017ApJ...848L..17C,2017ApJ...848L..18N} Further data of GRB 170817A have been in the mean time acquired in the X-rays and in the radio from $10^6$ s after the GRB trigger, and still ongoing.

It has been well established that short GRBs are produced by neutron star binary (NS-NS) mergers.\cite{1986ApJ...308L..43P,Eichler:1989jb,1992ApJ...395L..83N} Therefore, it is not surprising that GRB 170817A was labeled as such from the very beginning,\cite{2017ApJ...848L..12A,2017ApJ...848L..13A,2021ApJ...918...52L} despite the fact that it had been soon recognized that GRB 170817A was observationally very different from typical short GRBs.\cite{2017ApJ...848L..14G} Indeed, a comparison of GRB 170817A in the gamma-rays, X-rays and in the optical with typical short GRBs led \refcite{2018JCAP...10..006R1} to suggest that GRB 170817A looks more like a white dwarf binary (WD-WD) merger rather than an abnormal, special or unique NS-NS merger.\cite{2018ApJ...853L..10Y} identified additional sources similar to GRB 170817A and have proposed an alternative interpretation of them as WD-WD mergers.

In the mean time, $10^8$ s of data of GRB 170817A have been acquired in the X-rays, in the optical, and in the radio wavelengths, besides just the MeV radiation of the prompt emission. These observations have indeed led to alternative explanations. In fact:
\begin{itemize}

\item 
The NS-NS merger interpret the associated optical counterpart AT 2017gfo as a nuclear \textit{kilonova} produced by the decay of r-process yields in the matter ejected in the merger.\cite{2017Natur.551...67P,2017Natur.551...64A,2017ApJ...848L..17C,2017ApJ...848L..18N}

\item 
The experimental confirmation of the nuclear \textit{kilonova} needs a  univocal spectroscopic identification of the atomic species present in the ejecta.\cite{2021ApJ...918...10W,2021ApJ...910..116K,2021ApJ...906...94Z,2021ApJ...918...44B,2018ApJ...855...99C} This has not been achievable in view of lack of available accurate models of atomic spectra, the nuclear reaction network, density profile, and details of the radiative transport (opacity). Other mechanisms can also explain the photometric properties of AT 2017gfo, for instance the cooling of the expanding ejecta of a WD-WD merger.\cite{2018JCAP...10..006R1,2019JCAP...03..044R} We will further elaborate this scenario in this article.
\item 
The NS-NS merger leading to a jet propagating throughout the ejected matter appears in conflict with recent data by the \textit{Chandra} X-ray Telescope at $10^7$--$10^8$ s after the GRB trigger.\cite{2021arXiv210413378T,2020MNRAS.498.5643T}
\end{itemize}

In view of all the above, we here explore further, and extend, the suggestion by Ref. \refcite{2018JCAP...10..006R1} of GRB 170817A being the product of a WD-WD merger, adding new observations all the way up to $10^8$ s.
\begin{itemize}
    \item 
    The possibly observed re-brightening in the X-ray afterglow of GRB 170817A at 1000 days agrees with the predicted appearance of the pulsar-like activity of the newborn WD from a WD-WD merger.\cite{2018JCAP...10..006R1,2019JCAP...03..044R}
    \item 
    The rate of GRB 170817A-like events is well explained by the rate of WD-WD mergers.\cite{2018JCAP...10..006R1,2019JCAP...03..044R}
    \item 
    Interestingly, the host galaxy of GRB 170817, NGC 4993 distant at about 40 Mpc, is an old elliptical galaxy.\cite{2017ApJ...848L..12A} Elliptical old galaxies are amply recognized as preferred sites of type Ia supernovae produced by the so-called double-degenerate scenario, namely by WD-WD mergers.\cite{2018PhR...736....1L, 2020A&ARv..28....3D}
\end{itemize}

The aim of this article is to extend the treatment of Ref. \refcite{2019JCAP...03..044R} on WD-WD mergers, and exploit the analogy with the synchrotron emission in the X-rays, optical and radio bands in the afterglow of long GRBs\cite{2021MNRAS.504.5301R,2020ApJ...893..148R,2019ApJ...874...39W,2018ApJ...869..101R} to determine the emission of WD-WD mergers across the electromagnetic spectrum. Then, we apply the above considerations to the luminosity in the X-rays, optical and radio wavelengths observed in the afterglow of GRB 170817A.

We here show the prominent role of rotation and its effect on the synchrotron emission from the interaction of the newborn rotating object with the ejected matter in the merger. This process is energetically predominant and has been neglected in traditional simulations of these merging systems. The ejected matter expand in the magnetic field of the newborn fast rotating WD, which injects rotational and accretion energy into the expanding ejecta. While expanding, the ejecta radiate energy across the electromagnetic spectrum due to thermal cooling and synchrotron emission. We evidence that the newborn WD becomes observable as a pulsar when the synchrotron radiation fades off. The amount of mass ejected, the mass, rotation period, and strength of the magnetic field of the newborn WD are the most important features that determine the electromagnetic emission of the system.

We show that the above process leads to a hard-to-soft evolution of the emitted radiation with specific decreasing luminosities that approach a distinct power-law behavior. The late-time luminosity is dominated by the pulsar activity of the newborn object, therefore the asymptotic power-law gives information on the parameters of the newborn central object. The total energy radiated during the whole evolution is dominated by the energy injected and radiated from the central WD, so it is covered by its rotational energy.\cite{Li2018a} Energy and angular momentum conservation allow to infer, for instance, the spin and magnetic field of the newborn WD directly from the light-curve of the source, prior to any detailed fit of the observational data with the theoretical model (see Refs. \refcite{2020ApJ...893..148R,2019ApJ...874...39W,2018ApJ...869..101R}, for the case of long GRBs).

We apply the above considerations to GRB 170817A and show the agreement of the WD-WD merger scenario with all the available observational multiwavelength data from the gamma-rays all the way down to the radio wavelengths. The article is organized as follows. In Sec. \ref{sec:2}, we formulate the general physical conditions of the WD-WD coalescence that constrain the parameters of the newborn WD formed at merger. Section \ref{sec:3} presents an estimate of a possible mechanism leading to a gamma-ray prompt emission in these mergers, and how it compares with GRB 1780817A. Section \ref{sec:4} is devoted to the analysis of the  WD-WD post-merger early optical-infrared-ultraviolet emission by thermal cooling, and how it compares with AT 2017gfo. In Sec. \ref{sec:5}, we present the theoretical model of the synchrotron emission powered by the newborn WD, and how it leads to a multiwavelength emission (from the radio to the gamma-rays). A comparison with the emission of GRB 170817A $t\gtrsim 10^6$ s is presented. We outline our conclusions in Sec. \ref{sec:6}. We use cgs units throughout.

\section{Merging binary and post-merger remnant}\label{sec:2} 

The fate of the central remnant of a WD-WD merger with a total mass near (below or above) the Chandrasekhar mass limit can be one of three possibilities: (i) a stable newborn WD, (ii) a type Ia supernova, or (iii) a newborn neutron star. Sub-Chandrasekhar remnants can lead either to (i) and (ii), while super-Chandrasekhar remnants produce either (ii) or (iii). Super-Chandrasekhar remnants are supported by angular momentum, so they are less dense and metastable objects whose final fate is delayed until the excess of angular momentum is loss, e.g. via magnetic braking, inducing its compression.\cite{2019MNRAS.487..812B,2018ApJ...857..134B}

We are here interested in WD-WD mergers leading to stable, massive, sub-Chandrasekhar newborn WDs with a mass $\gtrsim 1.0 M_\odot$. These WDs can have rotation periods as short as $\sim 0.5$ s (see Ref. \refcite{2013ApJ...762..117B}) and can also avoid the trigger of unstable burning leading to type Ia supernova providing its central density is kept under some critical value of a few $10^9$ g cm$^{-3}$.\cite{2018ApJ...857..134B}

Numerical simulations of WD-WD mergers show that the merged configuration has in general three distinct regions:\cite{1990ApJ...348..647B,2004A&A...413..257G,2009A&A...500.1193L,2012A&A...542A.117L,2012ApJ...746...62R,2013ApJ...767..164Z,2014MNRAS.438...14D,2018ApJ...857..134B} a rigidly rotating, central WD, on top of which there is a hot, convective corona with differential rotation, surrounded by a rapidly rotating Keplerian disk. Roughly, half of the mass of the secondary star, which is totally disrupted, goes to the corona while the other half goes to the disk. The above implies that little mass is ejected in the merger. Numerical simulations show that the amount of expelled mass is approximated by\cite{2014MNRAS.438...14D}
\begin{equation}\label{eq:mej}
m_{\rm ej} \approx h(q)\,M, \quad 
h(q) = \frac{0.0001807}{-0.01672 + 0.2463 q - 0.6982 q^2 + q^3},
\end{equation}
where
\begin{equation}\label{eq:M}
    M = m_1 + m_2 = \left( \frac{1+q}{q} \right)m_2,
\end{equation}
is the total binary mass, and $q \equiv m_2/m_1 \leq 1$ is the binary mass ratio. Equations (\ref{eq:mej}) tells us that for a fixed total binary mass, the larger the mass symmetry, the smaller the mass that is ejected. Thus, for a fully symmetric mass ratio, $q=1$, the amount of expelled matter becomes $m_{\rm ej} \approx 3.4\times 10^{-4} M$.

WD-WD merger simulations show two important ingredients for our model. First, the central remnant (the newborn WD) is degenerate, namely massive ($\gtrsim 1.0 M_\odot)$, fast rotating, and magnetized.\cite{2018ApJ...857..134B} Second, although the amount of expelled matter is negligible with respect to the total mass of the system, the ejecta are crucial for the electromagnetic emission in the post-merger evolution.

We start with a double WD with components of mass $m_1$ and $m_2$, with corresponding radii $R_1$ and $R_2$. We shall make use of the analytic mass-radius relation\cite{1972ApJ...175..417N}
\begin{equation}\label{eq:MR}
    \frac{R_i}{R_\odot}= \frac{0.0225}{\bar{\mu}} \frac{\sqrt{1 - (m_i/M_{\rm crit})^{4/3}}}{(m_i/M_{\rm crit})^{1/3}},
\end{equation}
where $\bar{\mu}\approx 2$ is the molecular weight, and 
\begin{equation}\label{eq:Mcrit}
    M_{\rm crit} \approx \frac{5.816 M_\odot}{\bar{\mu}^2} \approx 1.4~M_\odot,
\end{equation}
is the critical mass of (carbon) WDs, $M_\odot$ and $R_\odot$ are the solar mass and radius. Since little mass is expelled, we estimate the newborn WD mass as
\begin{equation}\label{eq:mwdm2}
    m_{\rm wd} \approx M-m_d = m_1 + m_2 -m_d \approx \left(\frac{2 + q}{2\,q}\right)m_2,
\end{equation}
where we have approximated the disk mass by $m_d\approx m_2/2$, according to numerical simulations. Combining Eqs. (\ref{eq:M}) and (\ref{eq:mwdm2}), we obtain
\begin{equation}\label{eq:Mvsmwd}
    M \approx 2 \left(\frac{1 + q}{2+q}\right)m_{\rm wd},
\end{equation}
and using Eqs. (\ref{eq:mej}) and (\ref{eq:Mvsmwd}), we obtain
\begin{equation}\label{eq:mwdmej}
    m_{\rm wd} \approx \left(\frac{2 + q}{1+q}\right) \frac{m_{\rm ej}}{2\,h(q)}.
\end{equation}
As we shall see in Sec. \ref{sec:6}, the above equations allow us to infer, from the inferred mass of the ejecta from the fit of the multiwavelength data of GRB 170817A, the parameters of the merging components and of the newborn WD.

\section{The prompt $\gamma$-ray emission}\label{sec:3}

GRB 170817 was first detected by the gamma-ray burst monitor (GBM) on board the \textit{Fermi} satellite.\cite{2017ApJ...848L..14G} The gamma-ray emission was confirmed by \textit{INTEGRAL}.\cite{2017ApJ...848L..15S} 

\begin{figure}
    \centering
    \includegraphics[width=\hsize,clip]{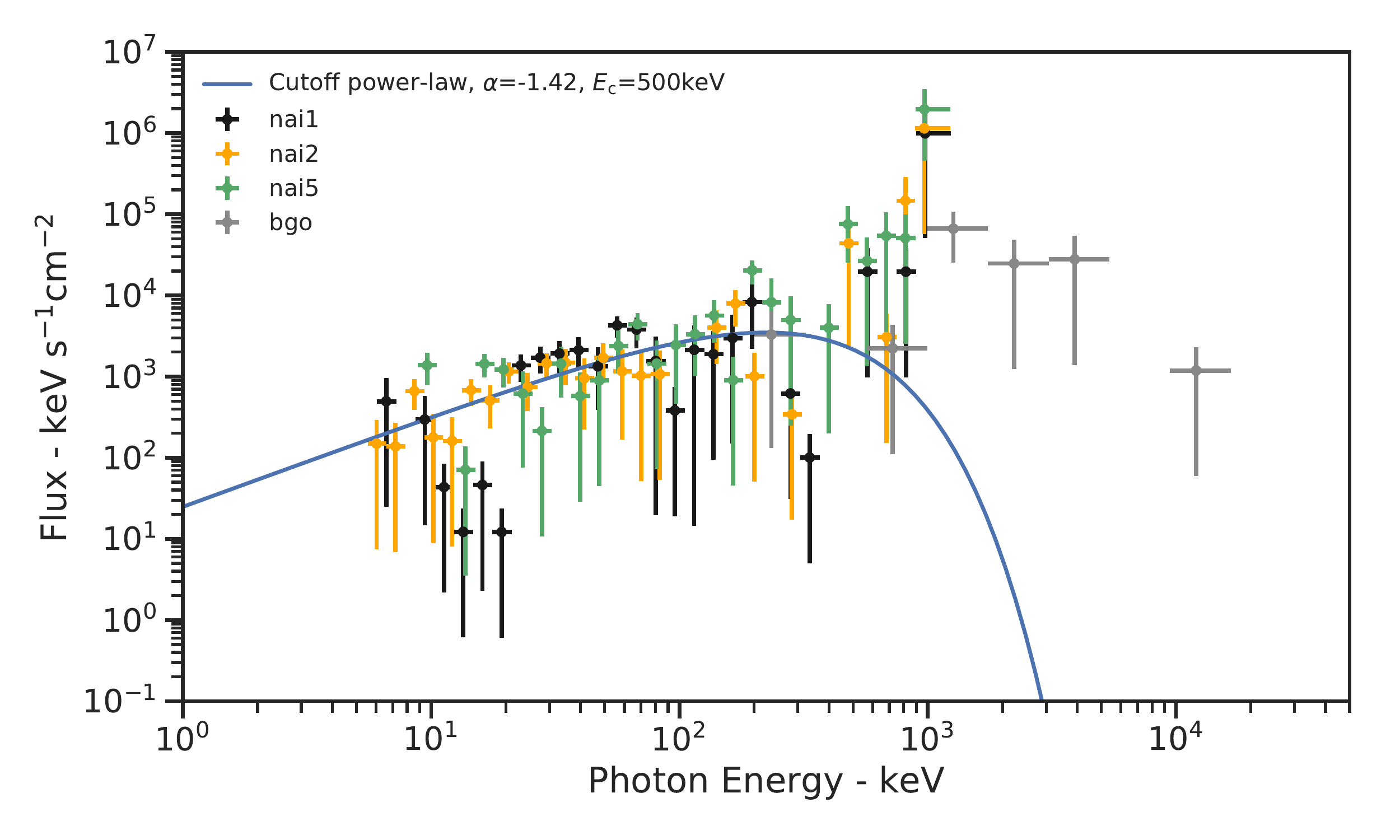}
    \caption{Spectral fits of $\nu F_\nu$ spectrum for the entire pulse ($-0.320$ to $1.984$ s) of the {\it Fermi}-GBM observation of GRB 170817A. This time interval is the best fit with a Comptonized function, with a cutoff energy $E_c=500\pm 317$ keV, $\alpha=-1.42\pm 0.18$, and time-averaged flux is $(1.84\pm 0.82)\times 10^{-7}$ erg s$^{-1}$ cm$^{-2}$ (see Sec. \ref{sec:3} for details of the data analysis).}
    \label{fig:LiangGBM}
\end{figure}

GRB 170817A is as a short burst with a duration ($T_{90}$) of $2.048$ s, as reported in the NASA/HEASARC database\footnote{https://heasarc.gsfc.nasa.gov/W3Browse/fermi/fermigbrst.html}. We performed a Bayesian spectral analysis of the \textit{Fermi}-GBM data by using the Multi-Mission Maximum Likelihood Framework (3ML, see Ref. \refcite{Vianello2015}), and the best model is selected by comparing the deviance information criterion (DIC, see Refs. \refcite{Spiegelhalter2002,Moreno2013}). We first fit the data with a single power-law function, and obtained a DIC value of $3138$. We then compare this model to the blackbody (Planck) spectrum over the same time interval, and obtained a DIC value of $3146$. We also fit the data with a Comptonized (i.e. a power-law with exponential cutoff, hereafter CPL) function, and obtained a DIC value of $3128$. The CPL model leads to a DIC improvement of $10$ with respect to the power-law model, and of $18$ with respect to the blackbody model, which suggests the CPL as the model that best fits the data. We refer to Refs. \refcite{Li2019a,Li2019c,Li2020,Li2021b,Li2021a} for a detailed Bayesian analysis of the data and the reduction procedure applied to GRBs.

As discussed above, the entire pulse ($-0.320$ to $1.984$ s) is best fitted by a CPL with a cutoff energy $E_c=500\pm 317$ keV and power-law index $\alpha=-1.42\pm 0.18$. The time-averaged flux is $(1.84\pm 0.82)\times 10^{-7}$ erg s$^{-1}$ cm$^{-2}$. With the measured cosmological redshift of $z=0.009783$, corresponding to a source distance of $\approx 43$ Mpc, the isotropic energy released in this time interval is estimated to be $(4.16^{+3.15}_{-1.84})\times 10^{46}$ erg. The non-thermal energy released at energies above $1$ MeV corresponds to only $2.82\%$ of the emission corresponding to $\approx 1.17\times 10^{45}$ erg. Therefore, most of the energy is released below MeV energies, which corresponds to $\approx 4.04\times 10^{46}$ erg. 

We here advance the possibility that the $\gamma$-ray prompt emission of GRB 170817A occurs from activity in the merged magnetosphere. We could think of the WD pulsar magnetosphere in an analogous way as the NS pulsar magnetosphere, therefore the presence of the strong magnetic field and rotation produces the presence of a electric field by Faraday (unipolar) induction.\cite{1969ApJ...157..869G} Numerical simulations show that the merger forms a transient hot corona with temperatures $10^8$--$10^9$ K that cools down rapidly mainly by neutrino emission.\cite{2018ApJ...857..134B} Therefore, thermal production of $e^+e^-$ pairs can occur for short time before it cools below the pair formation energy threshold. The charged particles are accelerated by the electric field to then follow the magnetic field lines generating both curvature and synchrotron photons. Since the magnetic field lines are curved, photon-photon collisions occur roughly in all directions, so the majority of the photons with energy in excess of $m_e c^2$ can decay into pairs again and generate a thermal plasma. A minority of photons escape along the rotation axis (see below), leading to the observed non-thermal emission above 1 MeV.

The cross-section of the $\gamma\gamma \to e^-e^+$ process is given by
\begin{equation}\label{eq:sigmagamma}
   \sigma_{\gamma \gamma} = \frac{3\sigma_T}{16} (1-\bar\beta^2) \left[ 2 \bar\beta (\bar\beta^2 - 2) + (3 - \bar\beta^4) \ln \left( \frac{1+\bar\beta}{1-\bar\beta} \right)\right],
\end{equation}
where $\sigma_T \approx 6.65\times 10^{-25}$ cm$^2$ is the Thomson cross-section, and $\bar\beta$ is the $e^-$ (or $e^+$) velocity (in units of $c$) in the center of momentum frame
\begin{equation}\label{eq:barbeta}
    \bar\beta \equiv \sqrt{1- \frac{2}{\bar\epsilon_{\rm inc} \bar\epsilon_{\rm tgt} (1-\cos\theta)}},
\end{equation}
being $\bar\epsilon_{\rm inc,tgt}\equiv\epsilon_{\rm inc,tgt}/(m_e c^2)$ the normalized energy of the incident and target photons which collide making an angle $\theta$ measured in laboratory frame.


Photons emitted along the curved magnetic field lines are expected to be absorbed since they will be radiated nearly isotropically. In this case, $\langle \cos\theta \rangle \sim 0$ and the cross-section becomes maximal at $    \bar\epsilon_{\rm inc} \bar\epsilon_{\rm tgt} \approx 4$, and $\sigma_{\gamma \gamma}\approx\sigma_T/4$. Under these conditions, the $\gamma\gamma$ optical depth is
\begin{equation}\label{eq:taugamma}
    \tau_{\gamma\gamma} \approx n_{\rm tgt} \sigma_{\gamma \gamma} r \approx \frac{L_{\rm tgt} \sigma_{\gamma \gamma}}{4\pi\,r\,c\,\epsilon_{\rm tgt}} \approx \frac{L_{\rm tgt} \,\bar\epsilon_{\rm inc}\,\sigma_T}{64\pi\,r\,m_e c^3},
\end{equation}
where $r$ is the source size and $n_{\rm tgt}$ is the density of target photons, which we have estimated as $n_{\rm tgt} \approx L_{\rm tgt}/(4\pi\,r^2\,c\,\epsilon_{\rm tgt})$, being $L_{\rm tgt}$ the luminosity emitted at energies larger than the target photon energy. 

For a transient hot corona, most photons are emitted at energies around the peak of the Planck spectrum, which for a temperature of a few $10^9$ K implies $\epsilon_{\rm inc}\sim \epsilon_{\rm tgt} \sim 3 k T \sim 1$ MeV. Assuming a source size $r \sim R_{\rm wd} \sim 10^9$ cm, and a target luminosity $L_{\rm tgt} \sim 4 \pi R_{\rm wd}^2 \sigma T^4 \sim 10^{51}$ erg s$^{-1}$, the optical depth (\ref{eq:taugamma}) $\tau_{\gamma \gamma}\sim 10^{10}$.

The above conditions imply that most photons interact generating an optically thick pair plasma which explains the dominant blackbody component observed by \textit{Fermi}-GBM. The observed non-thermal component is explained if $\approx 1\%$ of the photons escape from the system, which can occur near the rotation axis of the WD. There, the interaction angle could approach values as small as $\cos\theta \sim 1$, thereby reducing drastically the photon-photon cross-section.

\section{Thermal cooling of the ejecta as origin of the kilonova}\label{sec:4}

The second observed emission associated with GRB 170817A is the optical counterpart at about $0.5$ d after the \textit{Fermi}-GBM trigger, i.e. AT 2017gfo.\cite{2017Natur.551...71T,2017ApJ...848L...6L,2017ApJ...848L..17C,2017ApJ...848L..18N} For the modeling of this thermal emission of the expanding ejecta, we must take into account that in a non-homogeneous distribution of matter, the layers reach transparency at different times. For simplicity, we consider the ejected matter as a spherically symmetric distribution extending at radii $r_i \in [R_*,R_{\rm max}]$, with corresponding velocities $v_i \in [v_*,v_{\rm max}]$, in self-similar expansion
\begin{equation}\label{eq:radius}
    r_i(t) = r_{i,0} \hat{t}^n,\quad v_i(t) = n \frac{r_i(t)}{t} = v_{i,0} \hat{t}^{n-1},
\end{equation}
where $\hat{t} \equiv t/t_*$, being $t_* \equiv n R_{*,0}/v_{*,0}$ the characteristic expansion timescale, which is the same for all layers in view of the condition of self-similarity. Here, $r_{i,0}$ and $v_{i,0}$ are the initial radius and velocity of the layer (so at times $t \ll t_*$ close to the beginning of the expansion. The case $n=1$ corresponds to a uniform expansion.

The density at the position $r=r_i$ is given by
\begin{equation}\label{eq:rhoej}
\rho(r_i) = \frac{(3-m)}{4\pi} \frac{m_{\rm ej}}{R_{*,0}^3}\left[ \left(\frac{R_{\rm max}}{R_*}\right)^{3-m} -1 \right]^{-1}\left( \frac{r_i}{R_*} \right)^{-m} \hat{t}^{-3 n},
\end{equation}
where $m_{\rm ej}$ is the total mass of the ejecta, and $m$ is a positive constant. The distribution and time evolution given by Eq. (\ref{eq:rhoej}) ensures that at any time the total mass of the ejecta, i.e. the volume integral of the density, is always equal to $m_{\rm ej}$.

We divide the ejecta into $N$ shells defined by the $N+1$ radii 
\begin{equation}\label{eq:radii}
    r_{i,0} = R_{*,0} + i \frac{(R_{\rm max,0}-R_{*,0})}{N},\quad i=0,1,...,N,
\end{equation}
so the width and mass of each shell are, respectively, $\Delta r =  (R_{\rm max,0}-R_{*,0})/N$, and
\begin{equation}\label{eq:mi}
    m_i = \int_{r_i}^{r_{i+1}} 4\pi r^2 \rho(r) dr\approx \frac{4\pi}{m-3} r_i^2 \rho(r_i) \Delta r,
\end{equation}
so in view of the decreasing density with distance, the inner layers are more massive than the outer layers. The number of shells to be used must be chosen to satisfy the constraint that the sum of the shells mass gives the total ejecta mass, i.e.
\begin{equation}\label{eq:mjsum}
    \sum_{j=1}^{N} m_j = m_{\rm ej},
\end{equation}
where we have introduced the discrete index $j=i+1$ to differentiate the counting of the shells from the counting of radii given by Eq. (\ref{eq:radii}). In this work, we use $N=100$ shells which ensures that Eq. (\ref{eq:mjsum}) is satisfied with $99\%$ of accuracy. 

Under the assumption that the shells do not interact each other, we can estimate the evolution of the $i$-th shell from the energy conservation equation
\begin{equation}\label{eq:energybalance}
\dot{E}_i = -P_i\,\dot{V}_i - L_{{\rm cool},i} + H_{{\rm inj},i},
\end{equation}
where $V_{i} = (4\pi/3)r_{i}^3$, $E_i$, and $P_i$ are the volume, energy, and pressure of the shell, while $H_{{\rm inj},i}$is the power injected into the shell, and
\begin{equation}\label{eq:Lcoolabs}
     L_{{\rm cool},i} \approx \frac{c E_i}{r_i (1+\tau_{{\rm opt},i})},
\end{equation}
is the bolometric luminosity radiated by the shell, being $\tau_{\rm opt,i}$ the optical depth. 

Assuming a spatially constant, gray opacity throughout the ejecta, the optical depth of the radiation emitted by the $i$-th layer is given by
\begin{align}\label{eq:taui}
    \tau_{\rm opt,i} &= \int_{\infty}^{r_i} \kappa \rho(r) dr = \int_{R_{\rm max}}^{r_i} \kappa \rho(r) dr = \tau_{i,0} \hat{t}^{-2n} \\
    &\tau_{i,0} \equiv \frac{m-3}{m-1} \frac{\kappa m_{\rm ej}}{4\pi R_{*,0}^2}\frac{\left[ \left(\frac{R_*}{r_i}\right)^{m-1} - \left(\frac{R_*}{r_{\rm max}}\right)^{m-1} \right]}{\left[1- \left(\frac{R_*}{R_{\rm max}}\right)^{m-3}\right]},
\end{align}
where we have used Eq. (\ref{eq:rhoej}), and $\kappa$ is the opacity.

We adopt a radiation-dominated equation of state for the ejecta and, improving with respect to \refcite{2019JCAP...03..044R}, accounting for the radiation pressure, i.e.:
\begin{equation}\label{eq:ejectaEOS}
    E_i = 3 P_i\,V_i + L^{\rm abs}_{{\rm cool},i} \frac{r_i}{c}.
\end{equation}

The power injected into the ejecta originates from the newborn central WD.\cite{2019JCAP...03..044R} This energy is absorbed and thermalized becoming a heating source for the expanding matter. The power-law decreasing density (\ref{eq:rhoej} suggests that the inner the layer the more radiation it should absorb. In order to account for this effect, we weigh the heating source for each shell using the mass fraction, i.e.
\begin{equation}\label{eq:Hi}
    H_{{\rm inj},i} = \frac{m_i}{m_{\rm ej}} H_{\rm inj},
\end{equation}
where $m_i$ is the shell's mass, and adopt the following form for the heating source 
\begin{equation}\label{eq:Hinj}
    H_{\rm inj} = H_0 \left(1+\frac{t}{t_c}\right)^{-\delta},
\end{equation}
where $H_0$ and $\delta$ are model parameters. According to \refcite{2019JCAP...03..044R}, power from fallback accretion with $H_0 \sim 10^{45}$ erg s$^{-1}$, $\delta \sim 1.3$, and $t_c \sim t_*$ (see Table \ref{tab:parameters1}), dominates the energy release from the newborn WD at these early times.

The photospheric radius at a time $t$ is given by the position of the shell that reaches transparency at that time. Namely, it is given by the position of the shell whose optical depth fulfills $\tau_{\rm opt,i}[r_i(t)] = 1$. Using Eq. (\ref{eq:taui}), we obtain 
\begin{equation}\label{eq:Rph}
    R_{\rm ph} = \frac{R_{{\rm max},0} \hat{t}^n }{\left[1+\frac{m-1}{m-3}\frac{4\pi R_{*,0}^2}{\kappa m_{\rm ej}} \frac{ \left[1- \left(\frac{R_*}{R_{\rm max}}\right)^{m-3} \right]}{\left(\frac{R_{\rm max}}{R_*}\right)^{m-1}} \hat{t}^{2 n}\right]^{\frac{1}{m-1}}}.
\end{equation}
\begin{figure*}
    \centering
    \includegraphics[width=0.49\hsize,clip]{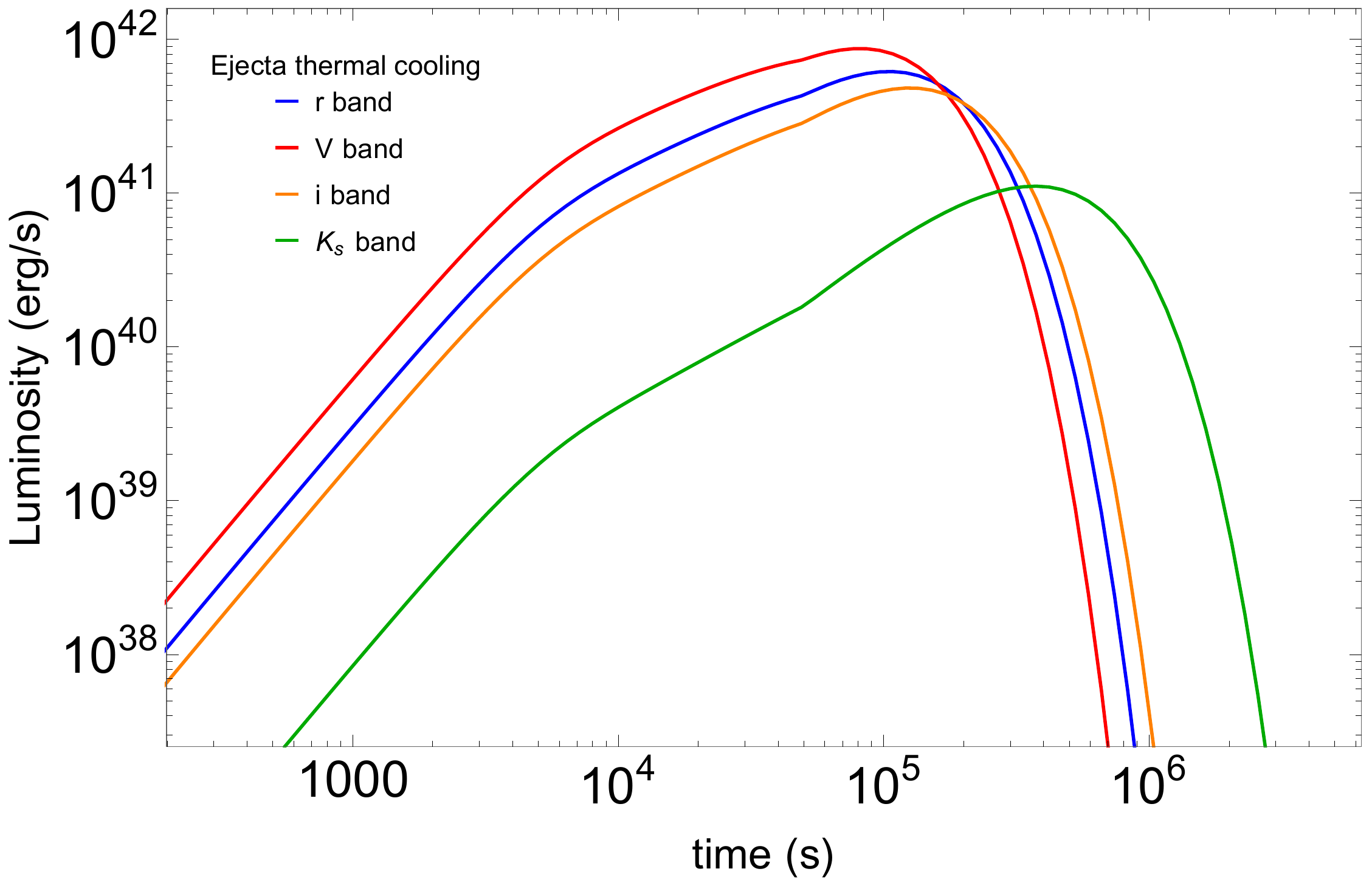}
    \includegraphics[width=0.49\hsize,clip]{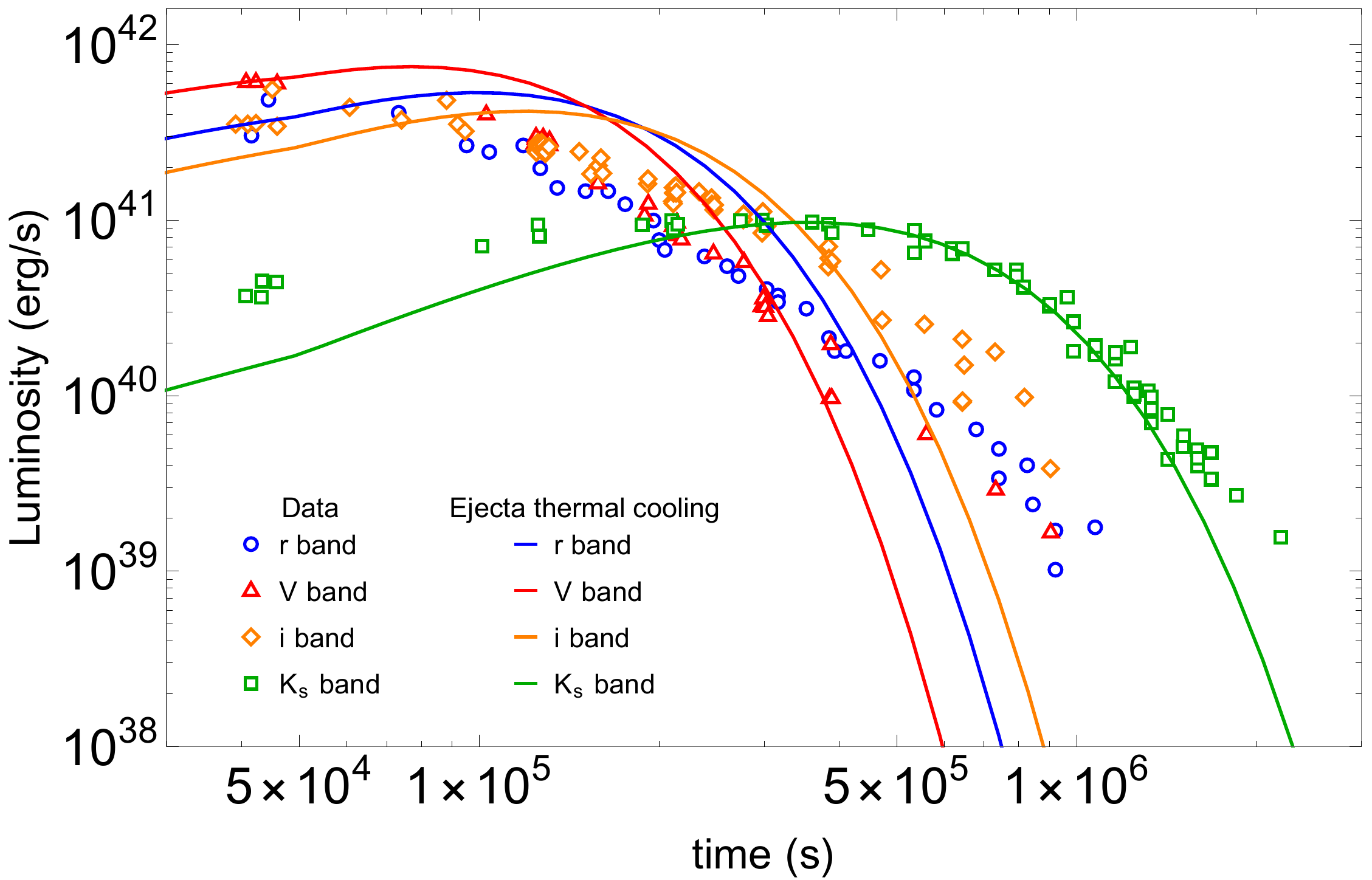}
    \caption{Left: emission from the expanding, cooling ejecta at early times in the visible (r and V) and in the infrared (i and $K_s$) bands, following the theoretical treatment of Section \ref{sec:6}. Right: zoomed view of the left panel figure at the times relevant for the comparison with the observational data of AT 2017gfo.\cite{2017Natur.551...71T,2017ApJ...848L...6L,2017ApJ...848L..17C,2017ApJ...848L..18N}}
    \label{fig:early_optical}
\end{figure*}

Equation (\ref{eq:Rph}) shows that, when the entire ejecta is optically thick, $R_{\rm ph} = R_{\rm max}$. Then, the transparency reaches the inner shells all the way to the instant over which $R_{\rm ph} = R_*$, reached at $t=t_{\rm tr,*}$, when the entire ejecta is transparent. The time $t_{\rm tr,*}$ is found from the condition $\tau_{\rm opt,*}[R_*(t_{\rm tr,*})] = 1$, and is given by
\begin{equation}\label{eq:ttr}
    \hat{t}_{\rm tr,*} = \left\{\frac{m-3}{m-1} \frac{\kappa m_{\rm ej}}{4\pi R_{*,0}^2}\left(\frac{R_*}{R_{\rm max}}\right)^{m-1}\frac{\left[ \left(\frac{R_{\rm max}}{R_*}\right)^{m-1} - 1 \right]}{\left[1- \left(\frac{R_*}{R_{\rm max}}\right)^{m-3} \right]}\right\}^{\frac{1}{2 n}}.
\end{equation}
At $t<t_{\rm tr,*}$, the photospheric radius evolves as $R_{\rm ph} \propto t^{\frac{n (m-3)}{m-1}}$, while at later times, $R_{\rm ph} \propto t^n$. For the parameters of our system, $t_{\rm tr,*}\sim 10^5$ s (see Fig. \ref{fig:early_optical}).

The bolometric luminosity is given by the sum of the luminosity of the shells
\begin{equation}
    L_{\rm bol} = \sum_{j=1}^{N} L_{{\rm cool},j},
\end{equation}
so the effective temperature of the thermal blackbody radiation, $T_s$, can be obtained from the Stefan–Boltzmann law, i.e. 
\begin{equation}\label{eq:Teff}
T_s = \left(\frac{L_{\rm bol}}{4\pi R_{\rm ph}^2 \sigma}\right)^{1/4},
\end{equation}
where $\sigma$ is the Stefan-Boltzmann constant. The power per unit frequency, per unit area, is given by the Planck's spectrum
\begin{equation}\label{eq:Bnu}
    B_\nu(\nu,t) = \frac{2\pi h \nu^3}{c^2} [e^{\frac{h\nu}{k_b T_s(t)}} - 1]^{-1},
\end{equation}
where $\nu$ is the radiation frequency, $h$ and $k_b$ are the Planck and Boltzmann constants. Most of the thermal cooling is radiated in the visible, infrared and ultraviolet wavelengths, which we refer to as optical. Therefore, the spectral density (power per unit frequency) given by the thermal cooling at a frequency $\nu$ is
\begin{equation}\label{eq:Lrad2}
   J_{\rm cool}(\nu,t) = 4\pi R_{\rm ph}^2(t) B_\nu(\nu,t),
\end{equation}
and the luminosity radiated in the frequency range $[\nu_1,\nu_2]$ can be then obtained as
\begin{equation}\label{eq:Lnucool}
    L_{\rm cool}(\nu_1,\nu_2; t) = \int_{\nu_1}^{\nu_2} J_{\rm cool}(\nu,t)d\nu.
\end{equation}

Figure \ref{fig:early_optical} shows the luminosity in the $r$, $V$, $i$, and $K_s$ energy bands obtained from Eq. (\ref{eq:Lnucool}), and compares them with the corresponding observations of AT 2017gf0. For the fit of this data, we have set the parameters as shown in Table \ref{tab:parameters1}. 

The value of the parameter $v_{\rm max,0}$ does not have any appreciable effect in the evolution, so it can not be constrained from the data. This happens because most of the mass is concentrated in the innermost layers, so they dominate the thermal evolution. For self-consistency of the model, we have set $v_{\rm max,0} = 2 v_{*,0}$, a value that keeps the outermost shell velocity well below the speed of light at any time in the evolution. As for the initial value of the internal energy of the shells, $E_i(t_0)$, we have set them to the initial kinetic energy of each layer, $E_i = (1/2) m_i v_i(t_0)^2$. 

\begin{table}
    \centering
    \begin{tabular}{l|r}
    Parameter & Value \\
    \hline
       $n$ & $1.22$\\
       $m$ & $9.00$\\
       $m_{\rm ej}$ ($10^{-3} M_\odot$) & $1.00$\\
       $R_{*,0}$ ($10^{11}$ cm) & $4.00$\\
       $v_{*,0}$ ($10^{9}$ cm s$^{-1}$) & $1.00$\\
       $\kappa$ (cm$^2$ g$^{-1}$) & $0.20$ \\
       $H_0$ ($10^{45}$ erg s$^{-1}$) & $8.16$ \\
       $\delta$ & $1.30$\\
       $t_c/t_*$ & $1.00$ \\
       \hline
    \end{tabular}
    \caption{Numerical values of the theoretical model parameters that determine the thermal cooling of the expanding ejecta which fits the data of AT 2017gf0 shown in Fig. \ref{fig:early_optical}.}
    \label{tab:parameters1}
\end{table}

There is a general agreement of the model with the observations, although it can not catch any detailed observational feature. There are some extensions to the present model that can increase its accuracy. For instance, we can abandon the assumption of spherical expansion allowing the layers to have a latitude-dependent velocity. Such a detailed treatment goes beyond our present scope that is to show the broad agreement of a WD-WD merger model with the multiwavelength data but not a dedicated model of AT 2017gfo.

\section{Synchrotron and WD pulsar radiation}\label{sec:5}

We have shown above that the expanding matter reaches full transparency at about $10^5$ s. After this time, the emission originated from the newborn WD as well as the one originated in the ejecta itself, become observable. We here follow the treatment in Ref. \refcite{2018ApJ...869..101R} for the explanation of the X-ray afterglow of long GRBs as originating from a newborn spinning NS powering the expanding SN. Here, we simulate the emission generated in the X-rays, in the optical, and in the radio by the synchrotron emission of electrons accelerated in the expanding magnetized ejecta, together with the emission of the newborn spinning WD pulsar.

We show below that synchrotron radiation originating in the merger ejecta dominates the emission up to nearly $10^8$ s. We find evidence of the newborn WD pulsar emission, owing to magnetic dipole braking, in the X-ray luminosity at approximately $10^6$ s, when the synchrotron radiation was not fully overwhelming yet, and then at times $10^8$ s, when the synchrotron luminosity sufficiently decreased for the WD pulsar emission to be fully observed (see Fig. \ref{fig:Lzooms} for details). 

\subsection{Synchrotron emission by the expanding ejecta}\label{sec:4dot1}

In this model, a fraction of the kinetic energy of the merger ejecta is used to accelerate electrons that, owing to the presence of the magnetized medium provided by the newborn WD, convert their kinetic energy into synchrotron radiation. The electrons are continuously injected from the newborn WD into the ejecta. The magnetic field threading every ejecta layer evolves as:
\begin{equation}\label{eq:B}
    B_i(t) = B_{i,0}\, \left[\frac{r_{i,0}}{r_i(t)}\right]^\mu = \frac{B_{i,0}}{\hat{t}^{\mu n}},
\end{equation}
where $B^{(0)}_i$ is the magnetic field strength at $r=r_{i,0}$, and $\mu$ gives the spatial dependence of the field at large distance from the newborn WD.

Because the electrons lose their energy very efficiently by synchrotron radiation (see details below), we can simplify our calculation by adopting that the radiation originates from the innermost layer of the ejecta, which we denote to as $R_*$. The evolution of this layer, following Eqs. (\ref{eq:radius}), is given by $R_*(t) = R_{*,0} \hat{t}^n$, $v_*(t) = v_{*,0} \hat{t}^{n-1}$, $t_* = n R_{*,0}/v_{*,0}$, and the magnetic field at its position decreases with time as $B_*(t) = B_{*,0} \hat{t}^{-\mu n}$ following Eq. (\ref{eq:B}).
 
The evolution of the distribution of radiating electrons is determined by the kinetic equation accounting for the particle energy losses\cite{1962SvA.....6..317K}
\begin{equation}\label{eq:kinetic}
    \frac{\partial N(E, t)}{\partial t}=-\frac{\partial}{\partial E}\left[\dot{E}\,N(E,t)\right] + Q(E,t),
\end{equation}
where $Q(E,t)$ is the number of injected electrons per unit time, per unit energy, and $\dot E$ is the electron energy loss rate. 

In our case, we assume electrons are subjected to adiabatic losses by expansion and synchrotron radiation losses, i.e. 
\begin{equation}\label{eq:gammadot}
    -\dot E = \frac{E}{\tau_{\rm exp}} + \beta B_*(t)^2 E^2,
\end{equation}
where $\beta = 2e^4/(3 m_e^4 c^7)$, $B(t)$ is the magnetic field, and
\begin{equation}\label{eq:tauexp}
    \tau_{\rm exp} \equiv \frac{R_*}{v_*} = \frac{t}{n} = \frac{t_*}{n}\hat{t},
\end{equation}
is the characteristic timescale of expansion. 

In order to find the solution to the kinetic equation (\ref{eq:kinetic}), we follow the treatment of \refcite{1973ApJ...186..249P}, adapted to our specific physical situation. We consider a distribution of the injected particles following a power-law behavior, i.e.
\begin{equation}\label{eq:Q}
Q(E,t)=Q_0(t)E^{-\gamma}, \qquad 0\leq E \leq E_{\rm max},
\end{equation}
where $\gamma$ and $E_{\rm max}$ are parameters to be determined from the observational data, and $Q_0(t)$ can be related to the power released by the newborn WD and injected into the ejecta. We assume that the injected power has the form
\begin{equation}\label{eq:Lt}
L_{\rm inj}(t) = 
L_0 \left(1+\frac{t}{t_q}\right)^{-k},
\end{equation}
where $L_0$, $t_q$, and $k$ are model parameters. We have not chosen arbitrarily the functional form of Eq. (\ref{eq:Lt}), actually, both the power released by magnetic dipole braking and by fallback accretion (see Eq. \ref{eq:Hinj}) obey this sort of time evolution.

Therefore, the function $Q_0(t)$ can be found from
\begin{equation}\label{eq:LandQ}
L_{\rm inj}(t) = \int_{0}^{E_{\rm max}} E\,Q(E,t) dE =  \int_{0}^{E_{\rm max}} Q_0(t) E^{1-\gamma} dE =Q_0(t)\frac{E_\mathrm{max}^{2-\gamma}}{2-\gamma},
\end{equation}
which using Eq.~(\ref{eq:Lt}) leads to
\begin{equation}\label{eq:Q0}
    Q_0(t) =
q_0\left(1+\frac{t}{t_q}\right)^{-k},
\end{equation}
where $q_0 \equiv  (2-\gamma)L_0/E_{\rm max}^{2-\gamma}$.

Having specified the evolution of the ejecta by Eq.~(\ref{eq:radius}) and the magnetic field by Eq.~(\ref{eq:B}), as well as the rate of particle injection given by Eqs.~(\ref{eq:Q}) and (\ref{eq:Q0}), we can now proceed to the integration of the kinetic equation (\ref{eq:kinetic}).

First, we find the evolution of a generic electron injected at time $t=t_i$ with energy $E_i$. Integration of Eq.~(\ref{eq:gammadot}) leads to the energy evolution
\begin{equation}\label{eq:gammavst}
    E = \frac{E_i\,(t_i/t)^n}{1 + {\cal M} E_i t^n_i\left[ \frac{1}{\hat{t}_i^{n (1+2\mu)-1}} -  \frac{1}{\hat{t}^{n (1+2\mu)-1}}\right]},
\end{equation}
where we have introduced the constant
\begin{equation}\label{eq:M2}
    {\cal M}\equiv \frac{\beta B^2_{*,0} t_*^{1-n}}{n (1 + 2 \mu) - 1},
\end{equation}
which have units of $1/({\rm energy}\times {\rm time}^n)$. In the limit $t/t_* \gg 1$ and $n=1$, Eq. (\ref{eq:gammavst}) reduces to Eq. (3.3) of Ref. \refcite{1973ApJ...186..249P}, and in the limit $t_* \to \infty$, reduces to the solution presented in Sec.~3 of Ref. \refcite{1962SvA.....6..317K} for synchrotron losses in a constant magnetic field.

The solution of Eq. (\ref{eq:kinetic}) is given by
\begin{equation}\label{eq:Nsol}
    N(E,t) = \int_E^\infty Q[E_i, t_i(t,E_i,E)] \frac{\partial t_i}{\partial E} dE_i,
\end{equation}
where the relation $t_i(t,E_i,E)$ is obtained from Eq. (\ref{eq:gammavst}).

We can write $N(E,t)$ as a piecewise function of time, separating it into different time intervals that allow simplifications and approximations depending upon the physical situation at work, and on the behavior of the energy injection given by Eq. (\ref{eq:Q0}). All the observational data of GRB 170817A is contained in the time interval $t<t_b$ and at electron energies int he range $E_b<E<E_{\rm max}$ (see definition of $t_b$ and $E_b$ below) where synchrotron losses are dominant. Under these conditions, the solution of Eq. (\ref{eq:Nsol}) is well approximated by
\begin{align}\label{eq:N3}
&N(E,t)\approx \begin{cases}
    \frac{q_0}{\beta B_{*,0}^2 (\gamma-1)}\hat{t}^{2 \mu n} E^{-(\gamma+1)}, & t < t_q\\
   \frac{q_0}{\beta B_{*,0}^2 (\gamma-1)}\left(\frac{t_q}{t_*}\right)^{k}\hat{t}^{2 \mu n-k} E^{-(\gamma+1)}, &   t_q<t<t_b,
\end{cases}
\end{align}
and we have defined
\begin{equation}\label{eq:Eb}
    E_b = \frac{\hat{t}^{2 \mu n-1}}{{\cal M} t_*^n},\quad
    t_b = t_* ({\cal M} t_*^n E_{\rm max})^{\frac{1}{2 \mu n-1}}.
\end{equation}

With the knowledge of $N(E,t)$, we can proceed to estimate the synchrotron spectral density (energy per unit time, per unit frequency) from $J_{\rm syn}(\nu,t)d\nu =  P_{\rm syn}(\nu,E) N(E,t) dE$, where $P_{\rm syn}(\nu,E)$ is the synchrotron power per unit frequency $\nu$, radiated by a single electron of energy $E$. Most of the synchrotron radiation is emitted in a narrow range of frequencies around the so-called photon critical frequency, $\nu_{\rm crit}$. Thus, we can assume electrons emit the synchrotron radiation at
\begin{equation}\label{eq:nuc}
    \nu \approx \nu_{\rm crit} \approx \alpha B E^2,
\end{equation}
where $\alpha = 3 e/(4\pi m_e^3 c^5)$. This gives a relation between the electron energy and the radiation frequency, and $P_{\rm syn}(\nu,E)$ can be approximated to the bolometric power
\begin{equation}\label{eq:Psynbol}
    P_{\rm syn}(\nu,E) \approx P_{\rm syn}(\nu) = \beta B^2_* E^2(\nu) = \frac{\beta}{\alpha} B_* \nu.
\end{equation}
Within this approximation, the spectral density is
\begin{equation}\label{eq:Jnu1}
    J_{\rm syn}(\nu,t) \approx P_{\rm syn}(\nu) N(E,t) \frac{dE}{d\nu}.
\end{equation}
It can be seen from Eq. (\ref{eq:N3}) that in each time and frequency interval we can write 
\begin{equation}\label{eq:Ngeneric}
    N(E,t) = \eta\,\hat{t}^l E^{-p},
\end{equation}
where $\eta$ and the power-law indexes $l$ and $p$ are known constants from Eq. (\ref{eq:N3}). With this, the spectral density (\ref{eq:Jnu1}) becomes
\begin{equation}\label{eq:Jnu3}
    J_{\rm syn}(\nu,t) = \frac{\beta}{2} \alpha^{\frac{p-3}{2}} \eta B_{*,0}^{\frac{p+1}{2}}\hat{t}^{\frac{2 l-\mu n(p+1)}{2}}\nu^{\frac{1-p}{2}}.
\end{equation}

The synchrotron luminosity in the frequencies $[\nu_1,\nu_2]$ can be then obtained as
\begin{equation}\label{eq:Lnu}
    L_{\rm syn}(\nu_1,\nu_2; t) = \int_{\nu_1}^{\nu_2} J_{\rm syn}(\nu,t)d\nu,
\end{equation}
which in a narrow frequency band from $\nu_1=\nu$ to $\nu_2=\nu+\Delta\nu$ where $\Delta\nu/\nu\ll 1$, can be well approximated as
\begin{equation}\label{eq:Lnu2}
    L_{\rm syn}(\nu, t) \approx \nu J_{\rm syn}(\nu,t)= \frac{\beta}{2} \alpha^{\frac{p-3}{2}} \eta B_{*,0}^{\frac{p+1}{2}}\hat{t}^{\frac{2 l-\mu n(p+1)}{2}}\nu^{\frac{3-p}{2}},
\end{equation}
where we have used Eq. (\ref{eq:Jnu3}).
%

\subsection{WD evolution and pulsar emission}\label{sec:4dot2}

The central WD emits also pulsar-like radiation. We adopt a dipole+quadrupole magnetic field model.\cite{2015MNRAS.450..714P} In this model, the total luminosity of spindown is
\begin{align}\label{eq:Lsd}
    L_{\rm sd} &= L_{\rm dip} + L_{\rm quad} \nonumber \\
    &= \frac{2}{3 c^3} \Omega^4 B_{\rm dip}^2 R_{\rm wd}^6 \sin^2\chi_1 \left(1 + \xi^2 \frac{16}{45} \frac{R_{\rm wd}^2 \Omega^2}{c^2} \right),
\end{align}
where the parameter $\xi$ defines the quadrupole to dipole strength ratio as
\begin{equation}
    \xi \equiv \sqrt{\cos^2\chi_2+10\sin^2\chi_2} \frac{B_{\rm quad}}{B_{\rm dip}},
\label{eq:eta}
\end{equation}
and the modes can be separated: $\chi_1 = 0$ and any value of $\chi_2$ for the $m = 0$ mode, $(\chi_1, \chi_2) = (90^\circ, 0^\circ)$ for the $m = 1$ mode, and $(\chi_1, \chi_2) = (90^\circ, 90^\circ)$ for the $m = 2$ mode. 

The WD evolution is obtained from the energy balance equation
\begin{equation}\label{eq:Erot}
	-(\dot{W}+\dot{T}) = L_{\rm tot} = L_{\rm inj} + L_{\rm sd},
\end{equation}
where $W$ and $T$ are, respectively, the gravitational and rotational energy of the newborn WD. We can obtain an analytic, sufficiently accurate solution of Eq. (\ref{eq:Erot}) by noticing the following. The power injected in electrons $L_{\rm inj}$ is larger than $L_{\rm sd}$ and has a shorter timescale with respect to the spindown timescale (see Eq. \ref{eq:Lt} and Fig. \ref{fig:injection}), so at $t<t_q$, we have $L_{\rm tot} \approx L_{\rm inj}$.  At later times, $L_{\rm tot} \approx L_{\rm sd}$, so the luminosity should approach the spindown luminosity
\begin{equation}
    L_{\rm sd} = L_{\rm sd,0} \left( 1 + \frac{t}{\tau_{\rm sd}} \right)^{-s},
\end{equation}
where $s=(n_b+1)/(n_b-1)$, being $n_b$ the so-called braking index ($n_b=3$ for a pure dipole and $n_b=5$ for a pure quadrupole), and $\tau_{\rm sd}$ is the spindown timescale 
\begin{equation}\label{eq:tausd}
    \tau_{\rm sd} = \frac{1}{2 {\cal A} \Omega_0^2},
\end{equation}
being ${\cal A} = (2/3) (B_{\rm dip}^2 R_{\rm wd}^6)/(c^3 I)$, and $\Omega_0$ the initial angular velocity of the WD.

With the above, Eq. (\ref{eq:Erot}) is integrated analytically accounting for changes in the WD structure. We describe the WD as an effective Maclaurin spheroid,\cite{1969efe..book.....C} so the angular velocity, $\Omega$, is related to the spheroid eccentricity, $e$, by
\begin{equation}\label{eq:Omegaecc}
    \Omega^2 = 2 \pi G \rho g(e),\quad 
    g(e)=\frac{\left(3-2 e^2\right) (1-e^2)^{1/2}\arcsin (e)}{e^3}-\frac{3 \left(1-e^2\right)}{e^2},
\end{equation}
where $\rho = 3 m_{\rm wd}/(4\pi R_{\rm wd}^3)$ is the density of the sphere with the same volume of the spheroid, being $m_{\rm wd}$ and $R_{\rm wd}$ the corresponding values of the mass and radius of the WD. The total energy of the spheroid is also a function of the eccentricity as
\begin{equation}\label{eq:Etot}
    E = T + W = \pi G \rho I_0 {\cal F}(e),
\end{equation}
where $I_0 = (2/5) m_{\rm wd} R_{\rm wd}^2$, and 
\begin{equation}\label{eq:Fapp}
    {\cal F}(e) = -2 + \frac{3(1-e^2)^{2/3}}{e^2}+\frac{(4 e^2-3) (1-e^2)^{1/6}}{e^3}\arcsin(e)\approx -\frac{4 e^2}{15},
\end{equation}
being the last line a series expansion of the function ${\cal F}$ which is accurate enough for low values of the eccentricity and which allows to give an analytic solution for the eccentricity as a function of time.

Then, integrating Eq. (\ref{eq:Erot}) and using Eqs. (\ref{eq:Etot}) and (\ref{eq:Fapp}), we obtain
\begin{equation}\label{eq:evst}
e(t) \approx \sqrt{\frac{15 \Theta (t)}{4}},\quad \Omega \approx \sqrt{2 \pi G \rho \Theta (t)},
\end{equation}
where 
\begin{align}
    \Theta (t) &= -{\cal F}(e_0) + {\cal G}(t)\\
   {\cal G}(t)&= \frac{L_0 t_q}{\pi G \rho I_0 (k-1)} \left[\left(1+\frac{t}{t_q}\right)^{1-k}-1\right] + \frac{L_{\rm sd,0} \tau_{\rm sd}}{\pi G \rho I_0 (s-1)} \left[\left(1+\frac{t}{\tau_{\rm sd}}\right)^{1-s}-1\right],
\end{align}
being $e_0$ is the initial value of the spheroid eccentricity, and we have used that the function $g(e)$ in Eq. (\ref{eq:Omegaecc}) satisfies $g(e) = -{\cal F}(e)$, at the order of our approximation. We recall that the moment of inertia changes with the eccentricity as $I = I_0 (1-e^2)^{-1/3} \approx I_0(1+e^2/3)$. The corresponding parameters of the model that explains the afterglow emission at different wavelengths are presented in the next section.  

\section{Model parameters from the multiwavelength data}\label{sec:6}

\begin{table}
    \centering
    \begin{tabular}{l|r}
    Parameter & Value \\
    \hline
      $\gamma$ &  $1.13$\\
       $k$  & $2.70$\\
       $\mu$ & $1.50$\\
       $L_0$ ($10^{39}$ erg s$^{-1}$)& $1.80$\\
       $B_{*,0}$ ($10^{9}$ G) & $1.00$\\
       $E_{\rm max}$ ($10^6 \ m_e c^2$) & $1.00$\\
       $t_q$ ($10^7$ s) & $1.22$\\
       $\xi$ & $0.00$\\
       $B_{\rm dip}$ ($10^{10}$ G) & $1.30$ \\
       $P$ (s) & $12.21$\\
       \hline
    \end{tabular}
    \caption{Numerical values of the theoretical model of synchrotron radiation of Sec. \ref{sec:5} that fit the multiwavelength observational data of GRB 170817A as shown in Fig. \ref{fig:Lmulti}.}
    \label{tab:parameters}
\end{table}

\begin{figure}
    \centering
    \includegraphics[width=\hsize,clip]{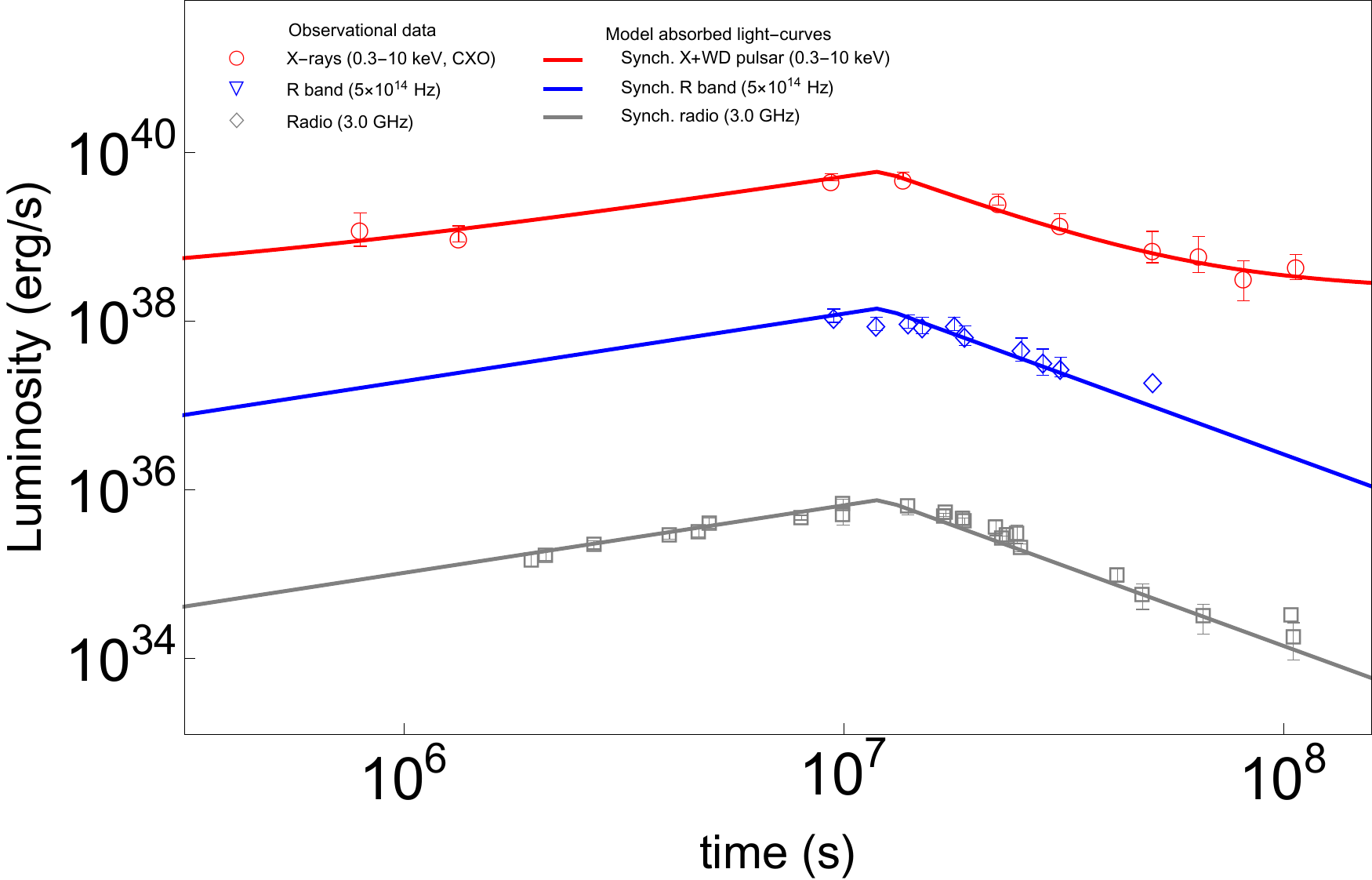}
    \caption{Comparison of the theoretical (solid curves) light-curves with the observational data (points) of GRB 170817A, in selected energy bands from the radio to the gamma-rays. The radio data at $3$ GHz have been taken from Refs.\cite{2017ApJ...848L..21A, 2017Sci...358.1579H, 2017ApJ...850L..21K, 2018Natur.554..207M, 2018ApJ...868L..11M, 2018ApJ...863L..18A, 2018ApJ...858L..15D, 2019MNRAS.483.1912P}; the infrared (F606W HST band) data points are retrieved from Refs.\cite{2018NatAs...2..751L, 2019ApJ...870L..15L, 2019MNRAS.483.1912P}; the X-ray ($0.3$--$10$ keV) data from CXO are taken from Ref.\cite{2021arXiv210402070H}.}
    \label{fig:Lmulti}
\end{figure}

We proceed to determine the model parameters that best fit the GRB 170817A afterglow. We list in Table \ref{tab:parameters} the value adopted for each parameter of the present model to fit the multiwavelength data of GRB 170817A shown in Fig. \ref{fig:Lmulti}. We did not consider here data at MeV energies because it is only present in the prompt emission that we have already discussed in Sec. \ref{sec:3} and is explained by a different mechanism from the synchrotron radiation. There are observations in the $30$ MeV-$10$ GeV energy band by AGILE \cite{2017ApJ...850L..27V} which give upper limits $\sim 10^{44}$--$10^{45}$ erg s$^{-1}$ in the time interval $\sim 10^3$--$10^6$ s. For the parameters of Table \ref{tab:parameters}, no emission is indeed expected at these energies because the maximum synchrotron radiation frequency obtained from Eq. (\ref{eq:nuc}) falls below $10$ GeV before $\sim 10^4$ s. The synchrotron luminosity vanishes at these energies at longer times.  

Having discussed the gamma-rays, we turn now to the X-rays, optical and radio emission. Figure~\ref{fig:Lmulti} compares the absorbed luminosity predicted by the model (see Sec. \ref{sec:4}), as a function of time, in selected energy bands, with the corresponding observational data of GRB 170817A. We have here included the X-ray data the $0.3$--$10$ keV energy band from the \textit{Chandra} X-ray Observatory (CXO) including the latest observations,\cite{2021arXiv210402070H} the infrared data from the HST at $\approx 5\times 10^{14}$ Hz,\cite{2018NatAs...2..751L,2019ApJ...870L..15L,2019MNRAS.483.1912P} and the radio data at $3$ GHz.\cite{2017ApJ...848L..21A,2017Sci...358.1579H,2017ApJ...850L..21K,2018Natur.554..207M,2018ApJ...868L..11M,2018ApJ...863L..18A,2018ApJ...858L..15D,2019MNRAS.483.1912P, 2021arXiv210402070H}

\begin{figure*}
    \centering
    \includegraphics[width=0.32\hsize,clip]{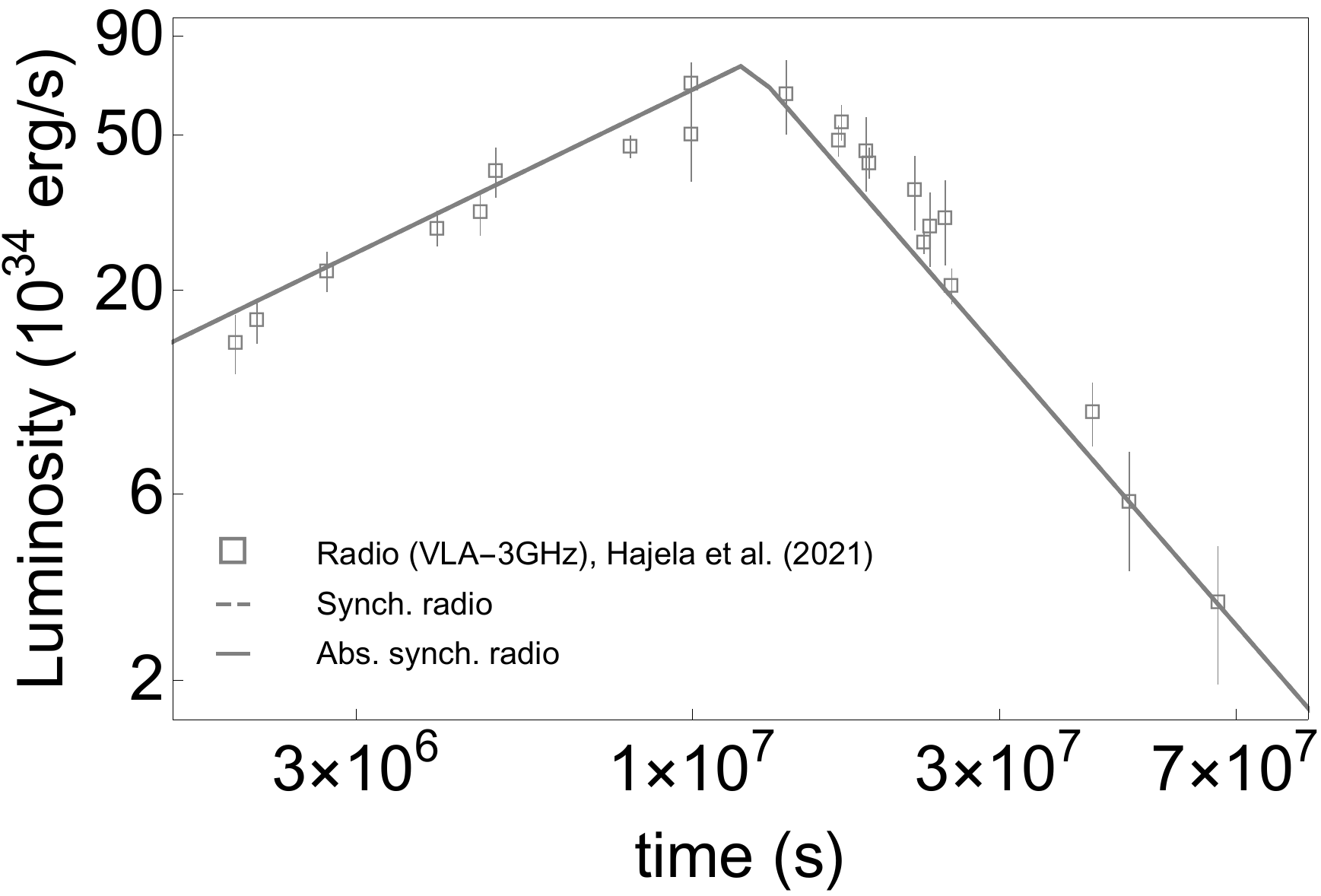}
    \includegraphics[width=0.32\hsize,clip]{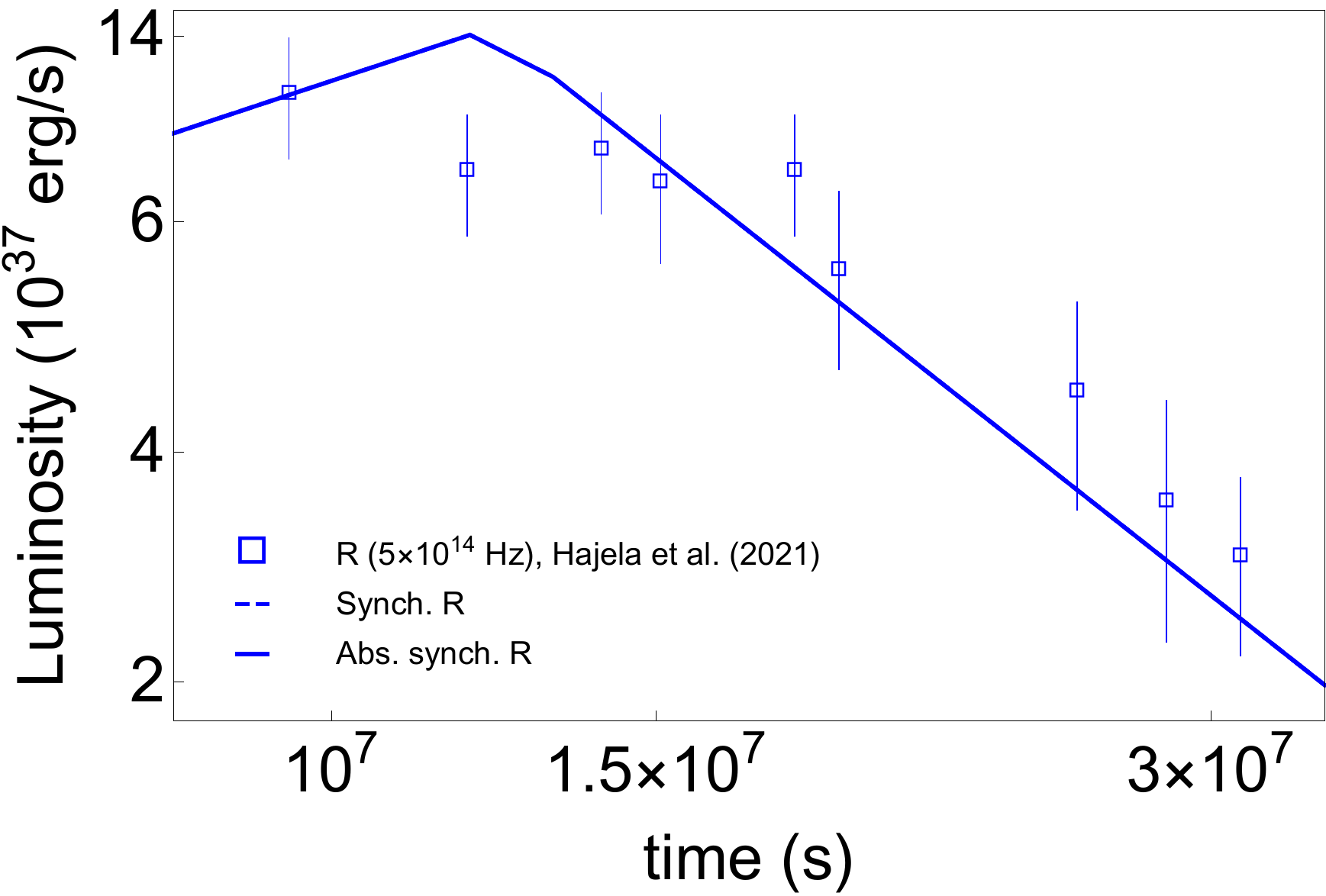}
    \includegraphics[width=0.32\hsize,clip]{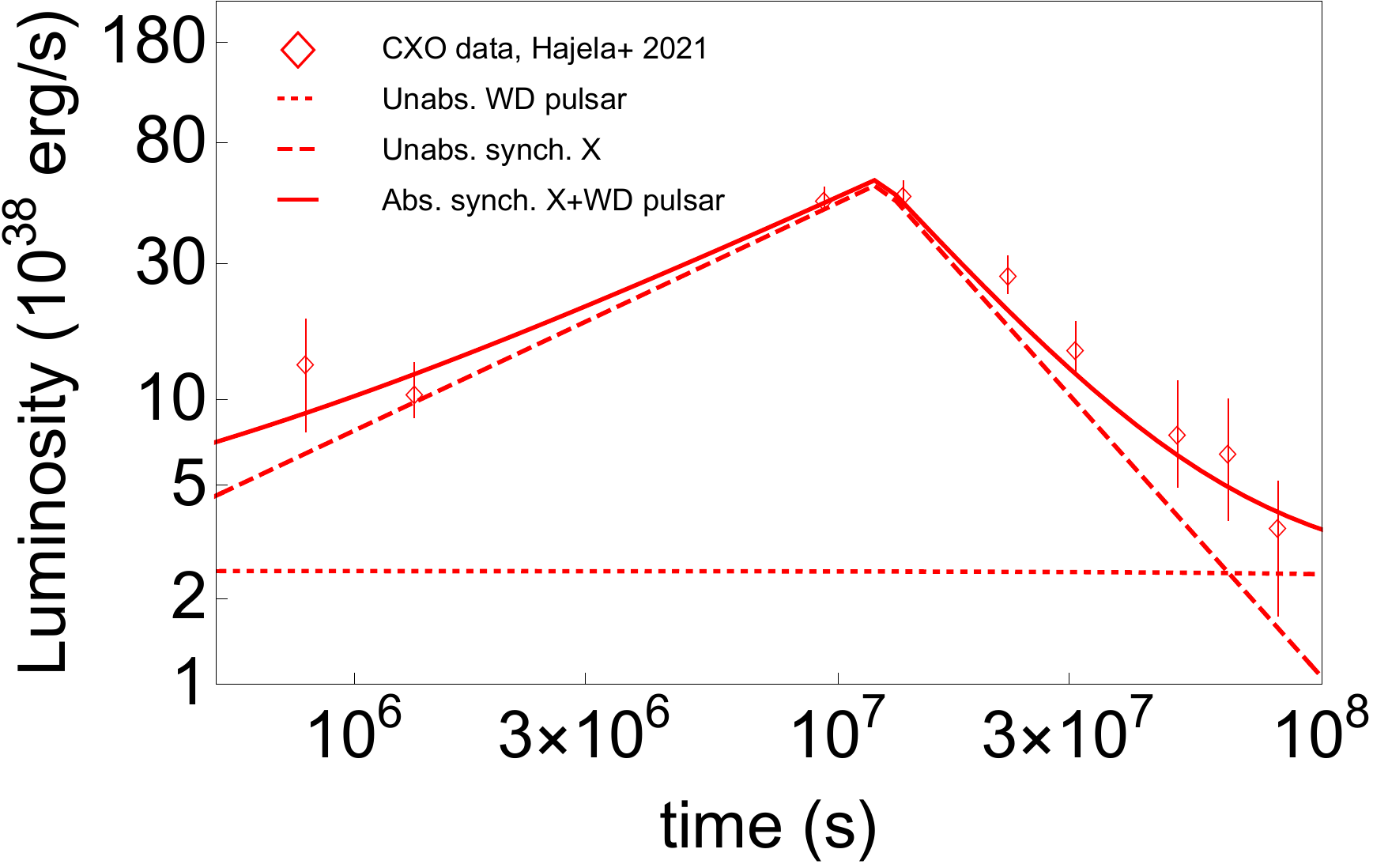}
    \caption{Zoomed views of the radio (left), optical (center), and X-ray (right) luminosities around transparency. The dashed curves represent the unabsorbed luminosities. The dotted curve in the right panel shows the contribution from the newborn WD pulsar which causes the deviation from the pure synchrotron power-law luminosity at times $\gtrsim 3\times 10^7$ s.}
    \label{fig:Lzooms}
\end{figure*}

The model shows a satisfactory fit of the data in the X-rays, optical and radio data, both where the luminosities rises, at times $t\sim 10^6$--$10^7$ s, and where it fades off, at $t \gtrsim 10^7$ s.  We show a closer view in Fig. \ref{fig:Lzooms} of the X-rays, optical, and radio luminosities around the time of the peak luminosity. The synchrotron luminosity rises as a power-law while the energy injection is constant, i.e. up to $t \approx t_q = 1.2 \times 10^7$ s, while it decreases as a power-law at later times. Probably the most interesting feature that can be seen from these zoomed views appears in the X-ray emission, where we can see in addition to the synchrotron luminosity, evidence of the WD pulsar emission owing to the magnetic dipole braking. The contribution from the pulsar emission is seen first at $t\sim 10^6$ s when the synchrotron radiation is rising but is still comparable with the pulsar spindown luminosity. Then, the synchrotron luminosity takes over, reaches a peak at approximately $10^7$ s, and then decreases. While the optical and radio counterparts continue to fade with time as dictated by the synchrotron radiation, the accuracy of the X-ray data of the CXO presented in \refcite{2021arXiv210402070H} allows to identify a clear deviation in the X-rays at a few $10^7$ s from such a power-law behavior. This is again the signature of the emergence of the WD pulsar emission.  

We have used the entity of this deviation to constrain the WD pulsar parameters. Since the pulsar emission depends on the WD radius (see Sec. \ref{sec:4dot2}), we first estimated the mass of the newborn WD. To accomplish this task, we must apply the considerations of Sec. \ref{sec:2}. From the inferred mass of the ejecta, $m_{\rm ej} = 10^{-3} M_\odot$ (see Table \ref{tab:parameters1}), we obtain an upper limit to the binary mass ratio via Eq. (\ref{eq:mwdmej}), by requesting that the newborn object be a stable, sub-Chandrasekhar WD, i.e. $m_{\rm wd} \lesssim 1.4 M_\odot$, which leads to $q\lesssim 0.87$. According to this maximum mass ratio and the ejecta mass value, Eq. (\ref{eq:mwdm2}) constraints the secondary mass to the range $m_2 \lesssim 0.85 M_\odot$. With the knowledge of $q$ and $m_2$, Eq. (\ref{eq:M}) constrains the total binary mass to $M\lesssim 1.82 M_\odot$. Thus, the primary component must satisfy $m_1 \lesssim 0.97 M_\odot$.

We assume that the newborn WD is stable, therefore it might have a mass close but not equal to the Chandrasekhar mass, since some mass will be accreted via matter fallback. Hereafter, we shall adopt in our estimates $m_{\rm wd} \approx 1.3 M_\odot$, so a radius $R_{\rm wd}\approx 3.4\times 10^8$ cm. With these WD structure parameters, we can proceed to constrain the magnetic field strength and rotation period.

The X-ray emission data shows that deviation from the pure synchrotron emission behavior starts at $\approx 3\times 10^7$ s, and extends up to when we have data, namely up to $\approx 10^8$ s (see Fig. \ref{fig:Lzooms}). This would suggest to chose this time for the spindown timescale $\tau_{\rm sd}$, but at the moment it is only a lower limit to $\tau_{\rm sd}$ because the luminosity did not reach yet the power-law given by the pulsar luminosity.

\begin{figure}
    \centering
    \includegraphics[width=0.8\hsize,clip]{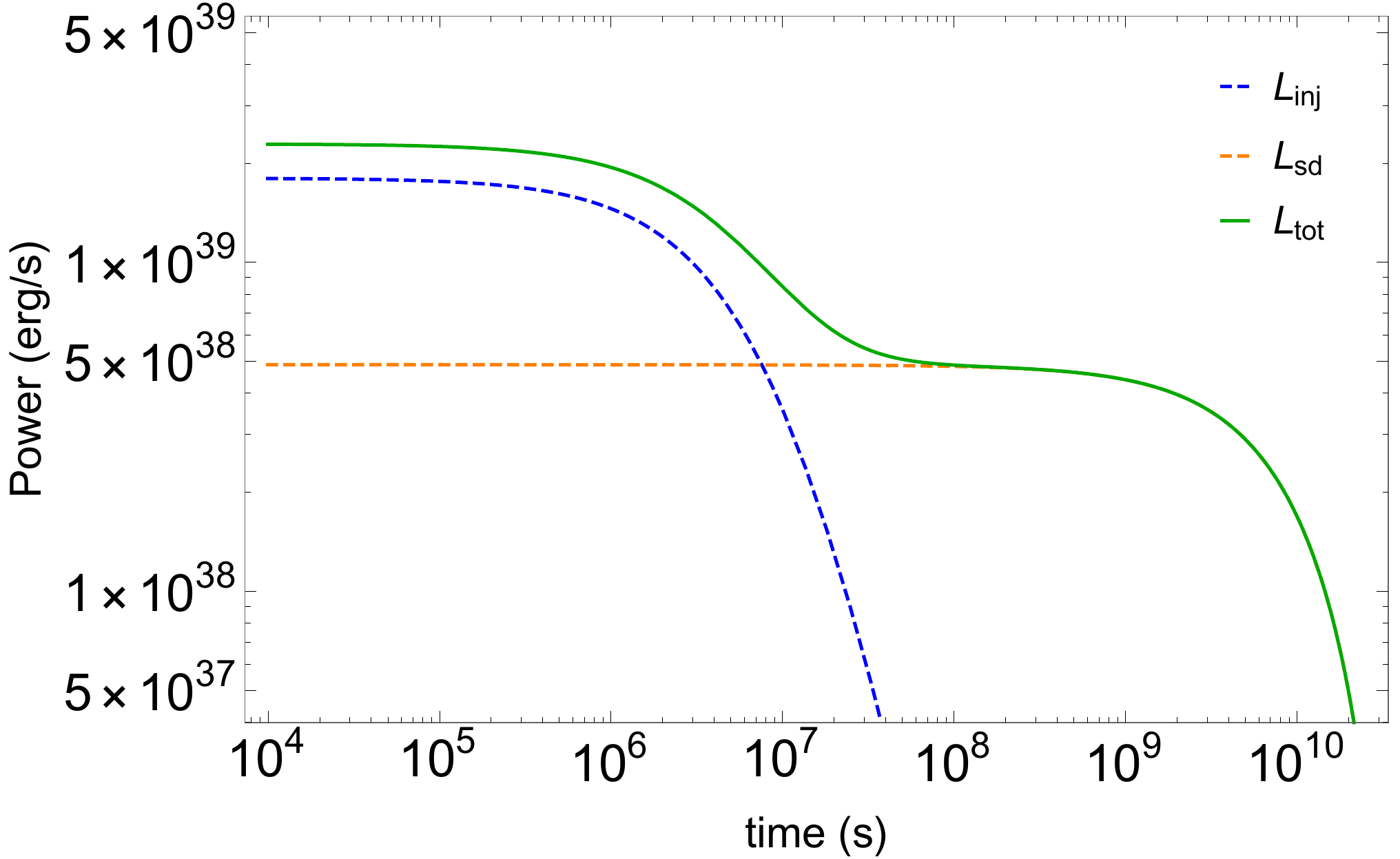}
    \caption{Power injected into the ejecta by the newborn WD and the pulsar emission as given, respectively, by Eqs. (\ref{eq:Lt}) and (\ref{eq:Lsd}). See the main text for further details.}
    \label{fig:injection}
\end{figure}

Hereafter, we assume a pure dipole (i.e. $\xi=0$) because the fit of the X-ray emission does not require at the moment the quadrupole component (see Fig. \ref{fig:Lzooms}). By eliminating the rotation angular velocity between the pulsar luminosity (\ref{eq:Lsd}) and the spindown timescale (\ref{eq:tausd}), we can express the magnetic field strength as
\begin{equation}\label{eq:Bvstau}
    B_{\rm dip}= \frac{3^{1/2} c^{3/2} I}{2^{3/2} L_{\rm sd}^{1/2} R_{\rm wd}^3 \tau_{\rm sd}},
\end{equation}
where $I$ is the moment of inertia. We can use Eq. (\ref{eq:Bvstau}) to give an upper limit to $B_{\rm dip}$ by setting as values of $L_{\rm sd}$ and $\tau_{\rm sd}$, the values of the latest value of the X-ray luminosity data, i.e. $L_{\rm sd} = L_X\approx 4.87 \times 10^{38}$ erg s$^{-1}$, and $\tau_{\rm sd}\approx 10^8$ s. With this, we obtain an upper value $B_{\rm dip, max} \approx 7.46\times 10^{11}$ G. To this upper value of $B_{\rm dip}$ it corresponds an upper value of the initial rotation period which can be obtained by calculating $P_0 = 2\pi/\Omega_0$ from Eq. (\ref{eq:Lsd}), i.e.
\begin{equation}\label{eq:Pmin}
    P_0 = 2\pi \left( \frac{2 B_{\rm dip}^2 R_{\rm wd}^6}{c^3 L_{\rm sd}} \right)^{1/4},
\end{equation}
from which we obtain $P_{0, \rm max} \approx 75.25$ s. We can further constrain the rotation period by seeking for values of the magnetic field strength and rotation period in agreement with the model presented in Sec. \ref{sec:3} for the prompt emission. Such a mechanism is expected to release magnetic energy stored in the magnetosphere, i.e. 
\begin{equation}\label{eq:Edip}
    E_B \approx \frac{1}{6} B_{\rm dip}^2 R_{\rm wd}^3,
\end{equation}
so we need a dipole magnetic field strength
\begin{equation}\label{eq:Bdip}
    B_{\rm dip} = \left(\frac{6 E_B}{R_{\rm wd}^3}\right)^{1/2} \approx \left(\frac{6 E_{\rm prompt}}{R_{\rm wd}^3}\right)^{1/2}.
\end{equation}
If we assume that the entire energy of the prompt emission, $E_{\rm prompt} \approx 4.16 \times 10^{46}$ erg (see Sec. \ref{sec:3}) is paid by the magnetosphere energy, we obtain a magnetic field $B_{\rm dip} = 9.61\times 10^{10}$ G. If we require the magnetic field energy to cover only the non-thermal component of the prompt, i.e. $1.17\times 10^{45}$ erg (see Sec. \ref{sec:3}), then the dipole magnetic field becomes $1.30\times 10^{10}$ G. For the above magnetic field values, Eq. (\ref{eq:Pmin}) gives, respectively, $P_0\approx 30$ s, and $P_0\approx 12$ s. The WD pulsar luminosity shown in Fig. \ref{fig:Lmulti} and \ref{fig:injection} corresponds to the latter case. 

The energy released (and injected into the ejecta) by the fallback accretion phase is $E_{\rm fb} = H_0 t_*/(\delta-1) \approx 3.34\times 10^{48}$ erg. Energy and angular momentum are transferred to the newborn WD during this phase, and since the rotational to gravitational energy ratio of a uniformly rotating WD is of the order of $10^{-2}$,\cite{2013ApJ...762..117B} the newborn WD has gained about a few $10^{46}$ erg of rotational energy during this phase. This might produce at a rotation period decrease of the order of a second, which confirms that the WD must be already fast rotating at birth. 

In Fig. \ref{fig:injection}, we plot the power injected in energetic electrons from the WD, $L_{\rm inj}$, and the luminosity due to magnetic dipole braking, $L_{\rm sd}$. Both components release an energy of the order of $10^{46}$ erg. From the inferred rotation period of $12.21$ s, the initial eccentricity turns out to be $e_0 \approx 0.39$, so the moment of inertia is about $5\%$ bigger than the one of the equivalent spherical configuration. Using the evolution equations (\ref{eq:evst}), we obtain that the moment of inertia, for instance from $10^4$ s to $10^8$ s, changes in about $0.03\%$. This small change in the structure of the WD, and the associated change in the rotational and gravitational energy, are sufficient to pay for the energy released by the ongoing magnetospheric phenomena responsible for the injection of particles into the ejecta and for the pulsar emission; see Sec. \ref{sec:4dot2}.

\section{Conclusions}\label{sec:8}

We have here addressed a self-consistent explanation of GRB 170817A, including its associated optical emission AT 2017gfo, based on a WD-WD merger. The most recent data of \textit{Chandra} of the X-ray emission of GRB 170817A at $\sim 10^8$ s ($\sim 1000$ d) after the GRB trigger,\cite{2021arXiv210413378T,2020MNRAS.498.5643T} indicates an X-ray re-brightening. This is explained by the emergence of the pulsar-like activity of the newborn WD (see Figs. \ref{fig:Lmulti} and \ref{fig:Lzooms}), as predicted in Refs. \refcite{2018JCAP...10..006R1,2019JCAP...03..044R}. We have here inferred that the newborn object is consistent with a massive ($\sim 1.3\,M_\odot$), fast rotating ($P\gtrsim 12$ s), highly magnetized ($B\sim 10^{10}$ G) WD, formed in a $1.0+0.8$ WD-WD merger (see Secs. \ref{sec:2} and \ref{sec:6}).

The post-merger emission at different wavelengths is explained as follows. The prompt gamma-ray emission detected by the \textit{Fermi}-GBM, with a luminosity of $\sim 10^{47}$ erg s$^{-1}$ and observed duration of $\lesssim 1$ s, can be explained by the transient hot corona produce at the merger. The high temperature produces photons that undergo $e^-e^+$ pair creation, the pairs are accelerated by the electric field induced by the $10^{10}$ G magnetic field an the WD rotation, thereby producing photons. The system is highly opaque to these photons (see Sec. \ref{sec:3}) to the $\gamma\gamma$ pair production process. Only a small percentage of photons is expected to be able to escape from the system along the polar axis, leading to the small amount of non-thermal emission observed above $1$ MeV, while the rest is expected to form a nearly thermal plasma.

The ejecta expand with velocities $\sim 10^9$ cm s$^{-1}$, and release energy by thermal cooling (see Sec. \ref{sec:4}) and synchrotron radiation (see Sec. \ref{sec:5}), powered by the newborn WD at the merger (see Sec. \ref{sec:4dot2}). Fallback accretion onto the newborn WD injects energy into the ejecta at early times, heating up the ejecta. The ejecta is optically thick up to nearly $10^5$, so the ejecta cool by diffusion while it expands. The thermal radiation is in agreement with the data of the early optical counterpart AT 2017gfo (see Fig. \ref{fig:early_optical}). This explanation is markedly different from the \textit{nuclear kilonova} from decay of r-process synthesized heavy nuclei in an NS-NS merger ejecta.

The signature of the synchrotron radiation is identified from nearly $10^6$ s, which explains the rising and decreasing luminosities with the same power-law slopes in the X-ray, optical and radio emissions (see Fig. \ref{fig:Lmulti} for details). 

The X-ray data is essential for identifying the emergence of the newborn WD as a pulsar. We have shown evidence of the pulsar emission around $10^6$ s and at late times $10^8$ s, causing the X-ray luminosity to deviate from the power-law emission of a pure synchrotron emission (see Figs. \ref{fig:Lmulti} and \ref{fig:Lzooms}). This data reveals a rotation period $\gtrsim 12$ s, and magnetic field of $\sim 10^{10}$ G. The follow-up of the GRB 170817A X-ray emission in the next months/years to come is crucial to confirm this prediction.

Summarizing, GRB 170817A/AT 2017gfo are explained by a WD-WD merger. The $10^{-3} M_\odot$ expelled in the merger expand and radiates via thermal and synchrotron cooling. The former explains AT 2017gfo and the latter the late-time X-rays, optical and radio emission. In this line, the association of GW170417A with GRB 170817A\cite{2017ApJ...848L..12A} is not confirmed in our treatment based on the new data in the X-rays, optical, and in the radio up to $10^8$ s. Therefore, we indicate the necessity to further inquire on the spacetime sequence of the early part of these events. 

Indeed, WDs of parameters approaching the present ones have been already identified, e.g. the WD in V1460 Her with $P\approx 39$ s,\cite{2021MNRAS.507.6132P} and the most recent observation of the WD in LAMOST J024048.51+195226.9 with $P\approx 25$ s.\cite{2021arXiv210811396P} WDs of similar properties have been proposed as a model of SGRs and AXPs.\cite{2017A&A...599A..87C,2013ApJ...772L..24R,2013A&A...555A.151B,2012PASJ...64...56M} Therefore, the newborn WD pulsar in GRB 170817A could show itself in the near future as an SGR/AXP in the GRB 170817A sky position. 



\end{document}